\documentclass[12pt,a4paper,final]{iopart}

\usepackage{iopams}

\usepackage{graphicx}
\usepackage{dcolumn}
\usepackage[usenames,dvipsnames,svgnames,table]{xcolor}
\usepackage[colorlinks=true,
            linkcolor=blue,
            urlcolor=black,
            citecolor=blue]{hyperref}
\usepackage{bm}
\usepackage{enumerate}
\usepackage{lipsum}
\usepackage{epstopdf}
\usepackage{float}
\usepackage[caption = false]{subfig}
\usepackage{tabularx}
\usepackage{color}

\begin{document}

\title{ Study of coupled-cluster correlations on electromagnetic transitions and hyperfine structure constants of W VI}

\author[cor1]{Anal Bhowmik$^{1}$, Sourav Roy$^2$, Narendra Nath  Dutta$^3$, Sonjoy Majumder$^1$}
\address{$^1$Department of Physics, Indian Institute of Technology Kharagpur, Kharagpur-721302, India.}

\address{$^2$School of Pharmacy, Hebrew University of Jerusalem, Jerusalem-9112001, Israel.}

\address{$^3$Department of Chemistry and Biochemistry, Auburn University,  Alabama-36849, USA.}

\eads{\mailto{analbhowmik@phy.iitkgp.ernet.in}}

\begin{abstract}
This work presents precise calculations of important electromagnetic transition amplitudes along with detail  of their many-body correlations using relativistic coupled-cluster method.  Studies of hyperfine interaction constants, useful for plasma diagnostic, with this correlation exhaustive many-body approach are another important area of this work. The calculated  oscillator strengths of allowed transitions, amplitudes of forbidden transitions and  lifetimes are compared with the other theoretical results wherever available and they show a good agreement.    Hyperfine constants of different isotopes of W VI,  presented in this paper will be helpful to get accurate picture of abundances of this element in different astronomical bodies. 

\end{abstract}

\maketitle

\section{INTRODUCTION}

Recent EBIT experiment \cite{Clementson2015} on extreme ultraviolet emission for few-times ionized tungsten motivates further study of the forbidden transitions in optical and near-infrared region as plasma diagnostic. The use of tungsten   as plasma facing materials attracts the interests of the scientists involved in tokamaks \cite{ Riccardo2009, Rohde2009, Reinke2010}. Though tungsten is available in nature as metal with body-center cubic structure, the presence of W$^{5+}$ in particular glasses \cite{Rakhimov2000} dictates the behavior of the glasses in magnetic field through the electron paramagnetic resonance (EPR) of the ion. Therefore, precise values of magnetic dipole transitions becomes important here. Recent work \cite{Mei2016} shows the importance of the study of hyperfine structure constant of this ion for the EPR estimation using crystal field theory.  There have been many  scientific studies on material containing neutral or ionized tungsten where detail spectroscopic data of these ions are required \cite{ Miyahara1991, Behrisch1991, Armstrong2013}. These spectroscopic data are mainly aimed towards ionization energies of low-lying fine structure, hyperfine levels, and various transition mechanisms among them.

Recent work of Safronova et al. \cite{Safronova2014} showed that long lived highly ionized atoms can be an excellent candidate for the frequency standard or to study $\alpha$ variation.     And the system, considered here, is ideal for infrared ion clock whose hyperfine levels requires to estimate precisely.  The first excited state of this ion, $5d_{5/2}$,  is a matastable state with the ground state,  $5d_{3/2}$. Therefore, the lifetime of this metastable state is important in order to use it as a fusion device in plasma medium \cite{Alfred}. All the forbidden lines among the low lying states of this ion are sensitive to the collisional de-excitations and present as indicators of electron density and temperature in the emission region in the study of astrophysics \cite{Seaton1954, Seaton1957} and laboratory tokamak plasmas \cite{Biemont1996}. To have precise excitation energy in plasma atmosphere, the estimations of hyperfine splittings, in other word hyperfine structure constants, of the system are indispensable. Due to the extremely high lifetimes of the few isotopes of this element, the spectroscopic study of tungsten and its highly stripped ions may take important role in the prediction of age and the procedure of formation and evolution of the astronomical bodies \cite{Lee2002, Masarik1997}. Moreover, the  technological developments of the high-resolution spectrometers increase the demand of the study of hyperfine structures of the various isotopes of this ion for different astrophysical purposes \cite{Casassus2005, Podpaly2009, Clementson2010}. Comparing with  highly accurate experimental hyperfine splitting results, our theoretical values are able to estimate  precise magnetic moment of a nucleus of W  with a non-zero nuclear spin. These moments affect any unpaired electrons associated with the atom which are useful to study the  EPR properties  of W VI  in molecules or cluster materials. \cite{Koutsospyros2006,Schmitz1992}.

The excitation energies \cite{Meijer1974, Sugar1975, Kramida2006, Loch2006, Kramida2009, Safronova2009,  Safronova2011,  Safronova2012a}, oscillator strengths \cite{Safronova2012b}, radiative rates \cite{   Safronova2009, Safronova2011, Safronova2012a, Seidel2009}, autoionization rates
 \cite{ Safronova2009, Safronova2011, Safronova2012a} , dielectronic satellite lines \cite{Safronova2009, Safronova2011, Safronova2012a, Safronova2012b}, and dielectronic recombination rates \cite{Safronova2009, Safronova2011, Safronova2012a, Safronova2012b, Seidel2009, Dalhed1986, Behar1997, Behar1999, Behar2000, Clementson2008, Meng2008,  Biedermann2009,     Meng2009, Radtke2009, Ballance2010, Schippers2011}  of various tungsten ions 
 have been studied in several recent literatures both theoretically and experimentally. Safronova \textit{et al.} \cite{Safronova2012b}  calculated some of the oscillator strengths of electric dipole transitions of W VI  by the relativistic all-order many-body perturbation theory using single and double excitations (SD) of the configuration space. Yoca \textit{et al.} \cite{Yoca2012}  and Migdalek \textit{et al.} \cite{Migdalek2014} calculated the   oscillator 
 strengths of a number of transitions of this ion using core polarization augmented relativistic Hartree-Fock method, which they named as HFR+CPOL and DF+CP, respectively. All these results motivate us to reinvestigate those transitions. Many-body correlations
study is very important here as $3d$, $4d$ and $4f$ are core orbitals. Form our earlier papers \cite{Dixit2007,Dixit2008}, it is expected that we will get significant effect of pair correlation for these forbidden transitions.          Therefore, the discrepancies between the results  demand to have correlation exhaustive relativistic \textit{ab initio} 
 calculations for allowed and forbidden transitions  of W VI. Here we apply  highly correlated coupled-cluster theory \cite{Dixit2008, Dutta2016} on a relativistic platform (RCC) to calculate the various line parameters of allowed and forbidden transitions as well as the hyperfine structure constants of few low-lying energy levels for this ion. 

\section{THEORY}

 A brief introduction of the formalism of our calculation  using the coupled-cluster theory is discussed here and details of the mechanism are available in our  earlier publications \cite{Dixit2007, Dutta2016}. The coupled-cluster theory is one of the well-known many-body methods \cite{Urban1985, Bishop1987, Lindgren1987,   Raghavachari1989, Ilyabaev199293, Chaudhuri2003, Mani2009, Mani2011} that allows one to write the atomic or ionic wavefunction for a single valence system using the expression: 
\begin{equation}\label{1}
\vert \Psi_v \rangle=e^T \{1+S_v\}\vert \Phi_v \rangle.
\end{equation} 
Here we assume that the valence electron occupies the `$v$'th orbital of the atom or ion. $| \Phi_v \rangle= a_v^{\dagger} |\Phi_0\rangle$, where $|\Phi_v\rangle$ and $|\Phi_0\rangle$ are the DF wavefunctions for a single-valence open-shell and closed-shell systems, respectively.  The operators $T$ and $S_v$ produce single to multiple electron excitations with respect to the reference $|\Phi_0\rangle$ and $| \Phi_v \rangle$, respectively.  However, in this present case, we consider these excitations up to the level of single and double only. Some valence triple excitations are also included in the present formalism using a pertubative treatment \cite{Dixit2007}.  Such an approximation of the coupled-cluster theory to generate highly correlated wavefunctions is well-known as coupled-cluster with single, double and valence triple excitations (CCSD(T)) method and is well established as indicated in our earlier works \cite{Dixit2008, Dutta2011, Dutta2013, Roy2014, Dutta2014, Roy2015, Dutta2015, Bhowmik2017}.

The general matrix element of any arbitrary operator $\hat{O}$ can be expressed in the framework of the RCC theory as,

\begin{eqnarray}\label{2}
O_{k\rightarrow i}&=&\frac{{\langle \Psi_k \vert \hat{O} \vert  \Psi_i \rangle}}{\sqrt{\langle \Psi_k \vert \Psi_k \rangle \langle \Psi_i \vert \Psi_i \rangle
}}\nonumber\\
&=& \frac{{\langle \Phi_k \vert \{1+S_k^\dagger\}e^{T^\dagger}\hat{O} e^T \{1+S_i\}\vert  \Phi_i \rangle}}{\sqrt{{{\langle \Phi_k \vert \{1+S_k^\dagger\}e^{T^\dagger} e^T \{1+S_k\}\vert  \Phi_k \rangle}} {\langle \Phi_i \vert \{1+S_i^\dagger\}e^{T^\dagger} e^T \{1+S_i\}\vert \Phi_i \rangle}}} \nonumber\\
&=& \frac{1}{N}\left[\left \langle \Phi_k \left \vert \left\{\bar{O}+\left(\bar{O}S_{1i}+S^{\dagger}_{1k}\bar{O}\right)+\left(\bar{O}S_{2i}+S^{\dagger}_{2k}\bar{O}\right)+\cdots \right \}\right \vert \Phi_i \right \rangle \right]
\end{eqnarray}

Here, the factor $N$ accounts the normalization of the coupled-cluster wavefunctions. In the last equality of the expression ~\ref{2}, the difference of matrix elements corresponding to the operators $\bar{O}=e^{T^{\dagger}}\hat{O}e^{T}$ and $\hat{O}$ yields the contribution of core correlation. The lowest order Bruckner pair-correlation effect in these matrix elements is considered by the term $\bar{O}S_{1i}+S^{\dagger}_{1k}\bar{O}$. The core polarization effect is calculated from the term $\bar{O}S_{2i}+S^{\dagger}_{2k}\bar{O}$. Here the subscripts '1' and '2' indicate the single and double excitations, respectively. However, in addition to these, there are other higher-order coupled-cluster terms, like, $S^{\dagger}_{k}\bar{O}S_{i}+S^{\dagger}_{i}\bar{O}S_{k}$ and normalization correction $O_{k \rightarrow i}-NO_{k \rightarrow i}$ to a wavefunction which are included in the present theoretical approach. A detail explanation of the different correlation contributing factors is available in one of our recent paper \cite{Dutta2016}.

The strength of the present coupled-cluster method is that it can account electronic correlation to all orders in the perturbation theory \cite{Lindgren1985}.  Also, from a theoretical point of view, the presence of non-linear terms make this theory more correlation exhaustive at a particular level of excitation \cite{Dutta2016}. One of the limitation of the present method is the non-consideration of full triple excitations and other higher (quadruple and so on) excitations. But these contributions, in general, within the uncertainty of experimental error \cite{Pasteka2017}. The another drawback of the present approach is the truncations of the exponential factors in Eq.~\ref{2} to linear terms only $(e^T=1+T)$. This can be circumvent using normal coupled cluster method \cite{Bishop1991} which is beyond the scope of present paper.

The  expression of the oscillator strength and transition probabilities (in s$^{-1}$) for allowed ($E1$) and forbidden ($E2$ and $M1$) transitions are given in Ref. \cite{Dutta2011, Roy2014, Mondal2013}. The single-electron reduced matrix elements corresponding to the electric dipole , electric quadrupole , and magnetic dipole  transition operators are discussed in detail in many references. \cite{Dutta2013, Grant1974, Johnson1995}. The lifetime $\tau_k$ of a state $k$ can be calculated by considering all different channels of emissions to the states $i$ from  $k$,
\begin{equation}\label{3}
\tau_k= \frac{1}{\sum_i A_{k\rightarrow i}},
\end{equation}
where $A_{k\rightarrow i}$  represents the  probability of the transition from $k$ to $i$.

The hyperfine energy shift of an atom or ion is given by \cite{Dutta2015, Lindgren1985, Cheng1985}
\begin{equation}\label{4}
H_{{hfs}}\approx\frac{AK}{2}+\frac{1}{2}
\frac{3K(K+1)-4J(J+1)I(I+1)}{2I(2I-1)2J(2J-1)}B.
\end{equation}
Here $K=F(F+1)-I(I+1)-J(J+1)$. $A$ and $B$ are the two well-known hyperfine
structure constants \cite{Cheng1985}. The constant $A$ is associated with the magnetic dipole moment of the nucleus. The constant $B$ corresponds to the electric quadrupole moment of the nucleus. The mathematical expressions to calculate these constants for single valence systems (considering $v$-th 
orbital is the valence orbital with relativistic quantum number $\kappa_v$) are as follows \cite{Johnson1995}: 
\begin{eqnarray}\label{5}
A&=&\mu_N g_I\frac{\langle
J||\textbf{T}^{(1)}||J\rangle}{\sqrt{J(J+1)(2J+1)}} \nonumber \\
&=& -\frac{g_I\kappa_v}{j_v(j_v+1)}\langle v|\frac{1}{r^2}|v\rangle \times 13074.7  {MHz}
\end{eqnarray}
and
\begin{eqnarray}\label{6}
B&=&2eQ\sqrt{\frac{2J(2J-1)}{(2J+1)(2J+2)(2J+3)}}\langle
J||\textbf{T}^{(2)}||J\rangle,\nonumber \\
&=& Q \frac{2j_v-1}{2j_v+2}\langle v|\frac{1}{r^3} |v\rangle \times 234.965 {MHz}
\end{eqnarray}
where $\mu_N$ is the nuclear magneton, $g_I$ is the nuclear $g$-factor and $Q$ is the quadrupole moment of the nucleus.
$\textbf{T}^{(1)}$ and $\textbf{T}^{(2)}$ are the two operators which depend on the inverse radial powers of all the electronic coordinates \cite{Cheng1985}. Their single-particle reduced matrix element forms are discussed explicitly in Ref.\cite{Cheng1985}.

\section{RESULTS AND DISCUSSIONS}

For our calculations of different transitions and hyperfine properties, we consider the
 Fermi-type of nuclear charge distribution function \cite{Visscher1997}. The basis-set expansion
  technique \cite{Motecc1990} is used here to construct the single-particle DF
   orbitals, where each radial basis function is considered to have the Gaussian-type form.  The radial dependence of these Gaussian
functions are determined by optimizing two radial parameters $\alpha_0$ and  
$\beta$ \cite{Motecc1990}. In order to find these optimized parameters, we compare the results of expectation values of $\langle r \rangle $, $\langle 1/r \rangle$, and energies for the present DF orbitals as mentioned above with those corresponding quantities for the DF orbitals obtained using a sophisticated numerical approach in the GRASP92 code \cite{Parpia2006}.  This comparison leads to an extremely good agreement between the corresponding expectation values  at $\alpha_0=0.00525$ and 
$\beta=2.70$.  The number of Gaussian functions considered to generate the DF orbitals of $s$, $p$, $d$, $f$, $g$ and $h$ symmetries are  33, 30, 28, 25, 21 and 20, respectively. However,  due to the computational limit, the number of active DF orbitals for the RCC calculations are restricted  to  16, 15, 15, 14, 11, and 7,  respectively, from the lowest energies of the above mentioned symmetries.  Here the selection criteria of number of active DF orbitals employed in the RCC calculations was decided by the convergence of  correlation energy at the closed shell system.     In the following discussions, wherever the correlation contribution ($\delta_\mathsf{corr}$) is mentioned, it indicates the difference between the RCC and the corresponding DF results.
\begin{figure*}

{ \hspace{1.3cm} \includegraphics[width=120mm]{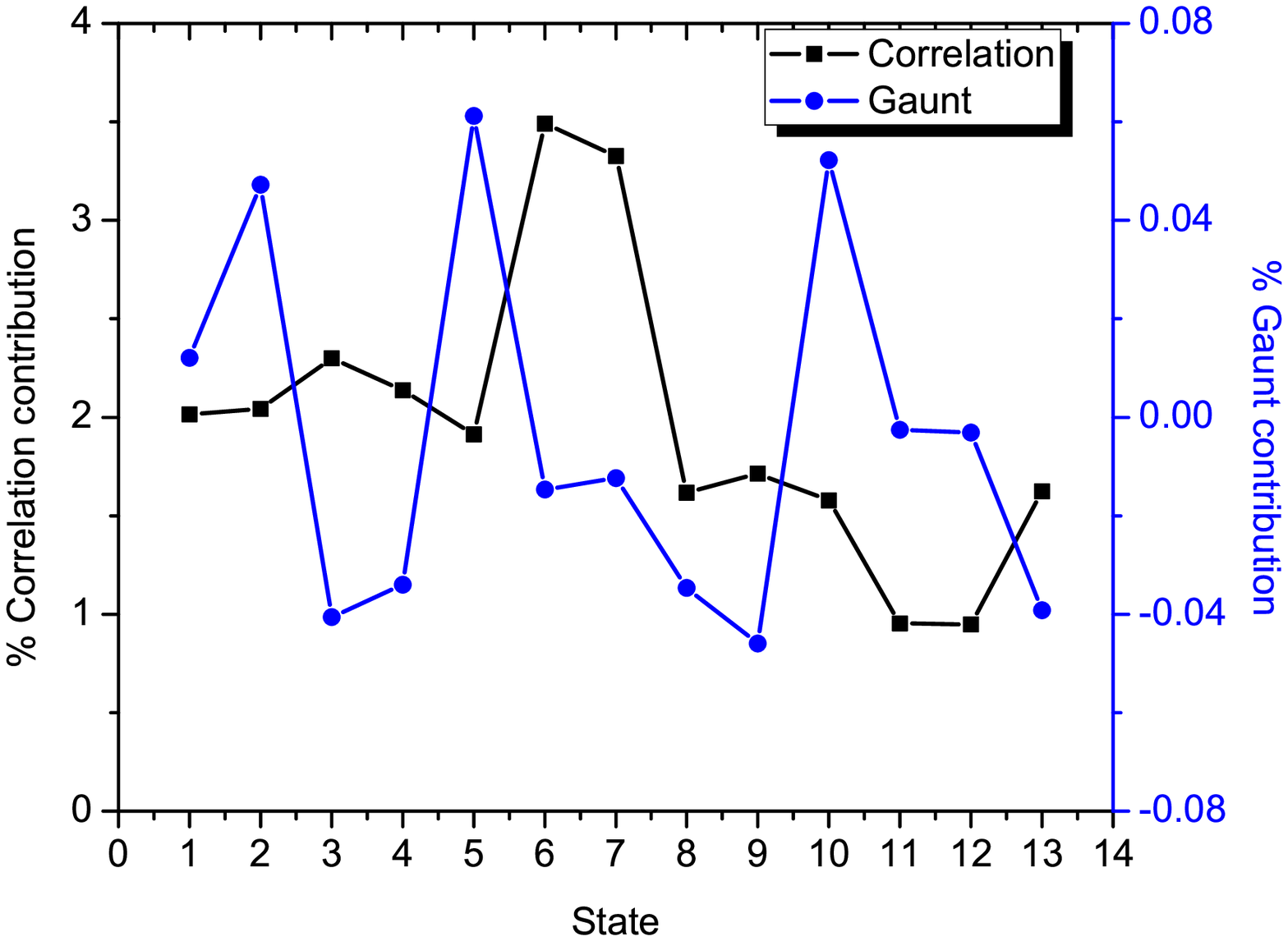}}\

\caption{Percentage of correlation and Gaunt contributions to the energy levels. Here numbers in the horizontal axis refer to the different energy states. They are 1$\rightarrow 5d_{3/2}$, 2$\rightarrow 5d_{5/2}$, 3$\rightarrow 6s_{1/2}$, 4$\rightarrow 6p_{1/2}$, 5$\rightarrow 6p_{3/2}$, 6$\rightarrow 5f_{5/2}$, 7$\rightarrow 5f_{7/2}$, 8$\rightarrow 7s_{1/2}$, 9$\rightarrow 7p_{1/2}$, 10$\rightarrow 7p_{3/2}$, 11$\rightarrow 5g_{7/2}$, 12$\rightarrow 5g_{9/2}$, 13$\rightarrow 8s_{1/2}$ .}
\label{A}
\end{figure*}

Fig. ~\ref{A} shows the Coulomb-correlation contributions to the ionization potentials  with respect to the DF results along with its relativistic effect (Gaunt interaction \cite{Gaunt1929, Mann1971}).  One can see from the figure that $5f_{5/2,7/2}$ states are maximally correlated (around 3.5\%). All the other  states have correlation contributions within 1.0\% to 2.3\%. Gaunt contributions to the ionization potential are too small, compared to the correlation contributions. It varies from -0.04\% to 0.06\% as shown in the figure.
 
 \begin{table}[!] 
\scriptsize
 \caption{ Calculated $E1$ transition amplitudes in length  gauge with the correlation  contributions (in a.u.) and velocity gauge results are kept for comparison. The experimental ($\lambda_{\textrm{NIST}} $)  and RCC ($\lambda_{\textrm{RCC}}$) wavelengths are presented   in \AA.}
\centering
\begin{tabular}{cccccccccccccc}

\hline \hline
  & &  & & &\multicolumn{5}{c}{Length gauge\hspace{0.1cm}} & & & \multicolumn{2}{c}{Velocity gauge}\\

\cline{6-10} 
   Transition           & &  $\lambda_{\textrm{NIST}} $   & $\lambda_{\textrm{RCC}}$&   & DF &	Core corr	&	Pair corr	&	Core pol
 & RCC &&& DF& RCC
        \\ [0.2ex]
\hline

$5d_{3/2}  $   $\rightarrow$ $5f_{5/2}$  &  & 382.1 & 383.4
&
 & 2.2130 &-0.0086	&	0.0119	&	-0.5229
  & 1.6776  &&&2.0858 & 1.6587\\
 
 \hspace{0.9cm}  $\rightarrow$ $6p_{1/2}$  &  & 677.7 &683.5
 &
 & -1.3542 &-0.0004	&	0.0326	&	0.1327
  & -1.1920  &&&-1.1568 & -1.0287\\ 
  \hspace{0.9cm}     $\rightarrow$ $6p_{3/2}$   & & 605.9 &609.6 &
 & 0.5427&	0.0007	&	-0.0153	&	-0.0410
  & 0.4886 &&&0.4651& 0.4224 \\ 
  \hspace{0.9cm}     $\rightarrow$ $7p_{1/2}$  &  &  &325.8 & & 0.3037&	0.0000	&	-0.0225	&	-0.0913
 & 0.1983 &&&0.2469&  0.1622\\ 
   \hspace{0.9cm}    $\rightarrow$ $7p_{3/2}$   & &  &317.1 & & 0.1408 &	0.0002	&	-0.0083	&	-0.0284& 0.1076  &&&0.1160&  0.0878 \\ 
  $5d_{5/2}  $    $\rightarrow$ $5f_{5/2}$  &  & 395.3 &395.4
 & & -0.6120 &	0.0025	&	-0.0082	&	0.1355  & -0.4770  &&&-0.5739 &  -0.4601 \\
    \hspace{0.9cm}   $\rightarrow$ $5f_{7/2}$   & & 394.1 &394.4
 & & 2.7283 &	-0.0113	&	0.0325	&	-0.5826& 2.1467  &&&2.5639&  2.0760\\ 
    \hspace{0.9cm}  $\rightarrow$ $6p_{3/2}$   & & 639.7 & 640.5
& & 1.7196 &	0.0013	&	-0.0240	&	-0.1212
& 1.5776  &&&1.4549& 1.3516 \\ 
    \hspace{0.9cm}  $\rightarrow$ $7p_{3/2}$   & &  &325.3 & & 0.4291 &	0.0002	&	-0.0302	&	-0.0825& 0.3251  &&&0.3462 & 0.2627\\ 
 $5f_{5/2}  $    $\rightarrow$ $5g_{7/2}$  &  & 994.6 &1004.3
 && 5.3059 &	-0.0008	&	-0.2423	&	-0.2684& 4.7550  &&&5.2420& 4.4892 \\ 
 $5f_{7/2}  $    $\rightarrow$ $5g_{7/2}$  &  & 1002.1 &1011.1
 && -1.0240 &	0.0002	&	0.0440	&	0.0514& -0.9218  &&&-1.0109&  -0.8595 \\
    \hspace{0.9cm}  $\rightarrow$ $5g_{9/2}$  &  & 1002.0 & 1011.0
&
 &  6.0599 &	-0.0010	&	-0.2604	&	-0.3042
& 5.4548 &&&5.9848 &  5.0864 \\
 $6s_{1/2}  $    $\rightarrow$ $6p_{1/2}$   & & 1468.0 &1438.7 && -1.9924 &	-0.0003	&	0.0386	&	0.2323& -1.7210 &&& -1.8953& -1.6888\\
    \hspace{0.9cm}  $\rightarrow$ $6p_{3/2}$  &  & 1168.1 &1146.2
 &
  & -2.7964 &	-0.0005	&	0.0560	&	0.3090
& -2.4312 &&&-2.6409&  -2.3606\\
    \hspace{0.9cm}  $\rightarrow$ $7p_{1/2}$  &  &  & 434.6 & & -0.1084 &	0.0001	&	-0.0142	&	-0.1092& -0.2182 &&&-0.1215&  -0.1968\\
    \hspace{0.9cm}  $\rightarrow$ $7p_{3/2}$  &  &  & 419.2 & & -0.0590 &	-0.0001	&	0.0169	&	0.1624
& 0.1032 &&&-0.0216&  0.0941\\ 
$7s_{1/2}  $    $\rightarrow$ $6p_{1/2}$   & &  761.3 &760.5
 && -1.0112 &	0.0001	&	0.0112	&	-0.0379& -1.0241  &&&-0.9791&  -0.9742\\
    \hspace{0.9cm}   $\rightarrow$ $6p_{3/2}$   & & 878.1 &879.1
 &
 & -1.7878 &	0.0002	&	0.0112	&	-0.0287
& -1.7860 &&&-1.7070&  -1.6924 \\ 
     \hspace{0.9cm}  $\rightarrow$ $7p_{1/2}$   & &  & 3436.1 & & -3.4689 &	0.0000	&	0.0564	&	0.1118 & -3.2932 &&&-3.4229 &  -3.2906 \\ 
    \hspace{0.9cm}  $\rightarrow$ $7p_{3/2}$  &  &  & 2662.1 & & 4.8206 &	0.0001	&	-0.0823	&	-0.1430
& 4.5853 &&& 4.6091 & 4.4304 \\ 
$8s_{1/2}  $    $\rightarrow$ $6p_{1/2}$   & &  &452.6 & & 0.3542 &	-0.0001	&	-0.0045	&	0.0245
 & 0.3711 &&&0.3403 &  0.3255 \\
 \hspace{0.9cm}   $\rightarrow$ $6p_{3/2}$   & &  &492.1 & & 0.5639 &	-0.0002	&	-0.0077	&	0.0160
 & 0.5692 &&&0.5283 &  0.5053 \\
   \hspace{0.9cm}  $\rightarrow$ $7p_{1/2}$  &  &  & 1656.6& & -1.8843 &	0.0000	&	0.0385	&	-0.0387
 & -1.8651 &&&-1.8813 &  -1.7889 \\ 
    \hspace{0.9cm}   $\rightarrow$ $7p_{3/2}$  &  &  &1926.6 & & 3.3418 &	0.0000	&	-0.0524	&	0.0380
& 3.3003 &&&3.2215
&  3.1125\\ 

 \hline 

\end{tabular}
\label{table:nonlin} 
\label{I}
\end{table}
 
Though we know core polarization contributes  majorly in electric dipole ($E1$) transition amplitudes,  in Table~\ref{I}, one can see that pair correlation contributions are significant.  Even in few cases, they are comparable to core polarization effect. This table shows our calculated reasults at the DF and the RCC levels  along with the different correlation contributing factors. The wavelengths are estimated from the excitation energies given by the National Institute of Standards and Technology (NIST)  website ($\lambda_{\textrm{NIST}} $) \cite{NIST2015} wherever available. The RCC wavelengths $(\lambda_{\textrm{RCC}})$ are also calculated theoretically using the present RCC method. The $E1$ amplitudes are calculated using both the length and  velocity gauge forms, and are found to have good agreement within 8\% in average.  It is known that the velocity gauge values are less stable compared to the corresponding length gauge values \cite{Grant2007}, and therefore, the latter gauge values are used commonly for the calculations of astrophysically important parameters, like, oscillator strengths, transition rates and lifetime.   The length gauge values for all the transitions are correlated at the level of 10\% or more than that, except the transitions associated with the $7s_{1/2}$ and $8s_{1/2}$ states.
Important correlation mechanism, core correlation (Core corr), core polarization (Core pol) and pair correlation (Pair corr) contributions to the total correlation in the different $E1$ transition amplitudes are also highlighted in the table. The difference between  $\delta_{{corr}}$ and the sum of the correlations from these three terms provides the correlation contribution from the sum of higher-order terms and normalization corrections to the wave functions as discussed in the theory section. Though our results support the conclusion of Yoca \textit{et al.} \cite{Yoca2012} that the core polarization contributions are the most important among the other correlation effects, the pair correlation contributions are significant and in some cases, like,  $5f_{5/2, 7/2}\rightarrow 5g_{7/2, 9/2}$ transitions, are comparable  with core polarization contributions. The figures in the parentheses show percentage values of theoretical uncertainties. These uncertainties are calculated by judging the quality of the wavefunctions of the states associated with a $E1$ transition where it peaks. Similar strategy  is taken for the uncertainty estimates of $E2$ transitions (Table~\ref{III}) and hyperfine $A$ constants (Table~\ref{VI}).   

\begin{table}[!]
  \caption{Weighted oscillator strengths of $E1$ transitions in length ($gf_\mathsf{L}$) and velocity ($gf_\mathsf{V}$) gauges (in a.u.)  and their comparisons with the other results (Others) .}
\centering
\begin{tabular}{cccc c c}

\hline \hline

   Transition     & $gf_\mathsf{L}$ & $gf_\mathsf{V}$ & & & Others
           \\ [0.2ex]
\hline

$5d_{3/2}  $  $\rightarrow$ $5f_{5/2}$  
& 2.237  & 2.187 & & & 2.008$^a$, 1.9055$^b$,1.800$^c$ \\ 
 \hspace{0.9cm}  $\rightarrow$ $6p_{1/2}$ 
 & 0.637  & 0.474 & & & 0.631$^a$, 0.6166$^b$,0.588$^c$ \\ 
 
   \hspace{0.9cm}  $\rightarrow$ $6p_{3/2}$   &   0.120 &  0.089 & & & 0.119$^a$, 0.1380$^b$, 0.116$^c$ \\ 
 
 \hspace{0.9cm}  $\rightarrow$ $7p_{1/2}$   
 & 0.037
  &0.025
 & & & 0.041$^c$\\   
 
 \hspace{0.9cm}   $\rightarrow$ $7p_{3/2}$ 
 & 0.011

  & 0.007

 & & & 0.013$^c$\\  

$5d_{5/2}  $    $\rightarrow$ $5f_{5/2}$  
 & 0.175  & 0.163 & & & 0.154$^a$, 0.1318$^b$, 0.138$^c$ \\ 
 
    \hspace{0.9cm}  $\rightarrow$ $5f_{7/2}$   &  3.552 &  3.321 & & & 3.195$^a$, 2.6303$^b$, 2.820$^c$ \\ 
   \hspace{0.9cm}   $\rightarrow$ $6p_{3/2}$   &  1.182 &  0.867 & & & 1.145$^a$, 1.1749$^b$, 1.092$^c$ \\ 
 
  \hspace{0.9cm}   $\rightarrow$ $7p_{3/2}$   
 & 0.099
  &0.064 & & &0.116$^c$ \\
 $5f_{5/2}  $    $\rightarrow$ $5g_{7/2}$   
& 6.905  & 6.155 & & & 6.439$^a$, 7.0795$^b$\\ 
 
  $5f_{7/2}  $    $\rightarrow$ $5g_{7/2}$   
& 0.258  & 0.224 & & & 0.241$^a$, 0.2630$^b$\\ 
 
   \hspace{0.9cm}   $\rightarrow$ $5g_{9/2}$   &  9.021 &  7.843 & & & 8.446$^a$, 9.1201$^b$\\

  $6s_{1/2}  $     $\rightarrow$ $6p_{1/2}$   &  0.613  & 0.590 & & & 0.606$^a$, 0.5888$^b$, 0.462$^c$ \\

    \hspace{0.9cm}  $\rightarrow$ $6p_{3/2}$  
& 1.537  & 1.449 & & & 1.521$^a$, 1.5136$^b$, 1.206$^c$ \\ 
 
   \hspace{0.9cm}   $\rightarrow$ $7p_{1/2}$  &  0.033
  & 0.027
 & & & 0.050$^c$\\  
 
  \hspace{0.9cm}   $\rightarrow$ $7p_{3/2}$  &  0.008
 & 0.006
 & & &0.012$^c$\\

  $7s_{1/2}  $    $\rightarrow$ $6p_{1/2}$   
& 0.418  & 0.379 & & & 0.398$^a$, 0.5370$^b$, 0.420$^c$ \\  
   \hspace{0.9cm}   $\rightarrow$ $6p_{3/2}$   &  1.103 &  0.991 & & & 1.045$^a$, 0.9332$^b$, 0.528$^c$ \\ 
\hline 

\label{table:nonlin} 
\label{II}

\begin{tiny}
\textbf{Note:} The wavelength corresponding to each transition can be found in Table~\ref{I}.
\end{tiny}\\
 
 \begin{tiny}
\hspace{-7.5cm} $a \rightarrow$ Ref. \cite{Safronova2012b},
 \end{tiny}\\

\begin{tiny}
\hspace{-7.5cm}$b \rightarrow$ Ref. \cite{Yoca2012},
\end{tiny}\\

\begin{tiny}
\hspace{-7.5cm}$c \rightarrow$ Ref. \cite{Migdalek2014}. 
\end{tiny}

\end{tabular}
\end{table}
 Table~\ref{II}  compares our oscillatior strength results  with the corresponding SD   \cite{Safronova2012b}, HFR+CPOL \cite{Yoca2012} and DF+CP values \cite{Migdalek2014}.  They all are falling in the vacuum ultra-violet (UV) region of the electromagnetic spectrum except the transitions $7s_{1/2} \rightarrow 7p_{1/2,3/2}$, which are in middle UV and near UV region. We calculate these oscillator strength values using  our calculated matrix elements and experimental wavelengths from the NIST  wherever available and so as the approaches by others whose calculations are available based on length gauge only. Therefore, differences among the oscillator strengths obtained from the various methods must come from the corresponding $E1$ amplitudes. This table shows an overall good agreement in the corresponding oscillator strength values as calculated by the RCC approach and as calculated using different other methods.

The significant pair correlation effect is obvious in Table~\ref{III}, where we present 
 $E2$ transition amplitudes  along with the corresponding transition wavelengths. These transitions are either falling in the  UV  or infra-red (IR) regions apart from the $6p_{1/2} \rightarrow 6p_{3/2}$ transition, which emits yellow light. There are a few  results for the $E2$ transitions in the literature obtained using the HFR+CPOL method \cite{Yoca2012} which are in  agreement with our RCC results. Unlike the $E1$ transitions, here pair correlation is very strong in many transitions and in some cases, like transitions from the $5f_{5/2,7/2}$ states,   this correlation factor contains lion share of the total correlation. The strong $E2$  transition amplitudes are estimated for the $5f_{5/2}   \rightarrow 6p_{1/2};7p_{1/2}$,  $5f_{7/2}   \rightarrow 6p_{3/2};7p_{3/2}$, $5g_{7/2}   \rightarrow 5g_{9/2}$, $6p_{1/2}   \rightarrow 6p_{3/2}$ and $7p_{1/2}   \rightarrow 7p_{3/2}$. High impact of correlations are observed for the $5d_{3/2}   \rightarrow 7s_{1/2};8s_{1/2}$, $5d_{5/2}  \rightarrow 7s_{1/2};8s_{1/2}$ transitions. These transitions are correlated by about  12\%, 30\%, 18\% and 28\%,  respectively. The correlation contributions to all the other transitions are less than 10 \%. 

\begin{table}[!]
\scriptsize
 
 \caption{Calculated $E2$ transition amplitudes with the different correlation  contributing terms (in a.u.).  The experimental ($\lambda_{\textrm{NIST}} $)  and RCC ($\lambda_{\textrm{RCC}}$) wavelengths are presented   in \AA. The `Other$^a$' indicates the results obtained from another method. }
\centering
\begin{tabular}{ c c cc c  c c c c c}

\hline \hline

   Transition      & $\lambda_{\textrm{NIST}} $&  $\lambda_{\textrm{RCC}}$   & DF  & Core corr  & Pair corr  & Core pol  &$\delta_\mathsf{corr}$   & RCC & Other$^a$
           \\ [0.2ex]
\hline

$5d_{3/2}  $    $\rightarrow$ $5d_{5/2}$  & 11482.4 & 12611.9

 & -1.7155  & 0.0103  & 0.0366 	& 0.0807 & 0.1231 & -1.5924  & -1.6610\\
 
    \hspace{0.9cm}  $\rightarrow$ $5g_{7/2}$   & 276.1&277.5
 & 3.0304  &-0.0032	 &-0.0700	&-0.0345  & -0.1078&  2.9226 &
\\ 
 
  \hspace{0.9cm}  $\rightarrow$ $6s_{1/2}$  &  1259.0&1302.0
 & 2.9061  &-0.0010 	&-0.0966	& -0.0525  & -0.1451&  2.7610 & 2.8519 \\ 
 
  \hspace{0.9cm}   $\rightarrow$ $7s_{1/2}$   & 358.5&360.0
 & 0.2351  & 0.0002  &	0.0143  & 0.0300   & 0.0272&  0.2623 &
\\ 
 
   \hspace{0.9cm}   $\rightarrow$ $8s_{1/2}$   & &272.3 &  -0.1015  &-0.0001 & -0.0131 &	-0.0265 & -0.0308
& -0.1323 &\\ 
 
$5d_{5/2}  $    $\rightarrow$ $5g_{7/2}$   & 282.9&283.7

& -1.0713  &0.0013	 &0.0036 	&0.0082 & 0.0133
& -1.0580 &\\
 
 \hspace{0.9cm}   $\rightarrow$ $5g_{9/2}$   & 282.9&283.7
 & 3.7912  &-0.0045 	&-0.0136 &	-0.0323 & -0.0517
& 3.7395 &\\ 
 
  \hspace{0.9cm}  $\rightarrow$ $6s_{1/2}$   & 1414.0& 1451.9
& 3.7129  &-0.0029	 &-0.0842 	&-0.0544 & -0.1390 & 3.5739 & 3.4949\\ 
 
  \hspace{0.9cm}   $\rightarrow$ $7s_{1/2}$   & 370.1& 370.6
& 0.3692  &0.0008 &0.0549	& 0.0314 & 0.0672 & 0.4364 & \\ 
 
  \hspace{0.9cm}   $\rightarrow$ $8s_{1/2}$   & &278.3 &  -0.1585  &-0.0005	& -0.0274 	&-0.0280 & -0.0447
& -0.2033 &\\ 
 
 $5f_{5/2}  $    $\rightarrow$ $5f_{7/2}$   & 133511.4 & 151745.1
 & -4.5606  &0.0020	& 0.3269	& 0.0473 & 0.4121
& -4.1484 &\\ 
 
   \hspace{0.9cm}    $\rightarrow$ $6p_{1/2}$   & 876.2& 873.3
& 9.0523  &0.0008 	&-0.4743	& -0.1071 & -0.6326&  8.4197 &
\\ 
 
  \hspace{0.9cm}  $\rightarrow$ $6p_{3/2}$   & 1034.7& 1033.5
& -5.1391  & -0.0010 & 0.2759& 	0.0558 & 0.3627&  -4.7765 &\\ 
  
   \hspace{0.9cm}   $\rightarrow$ $7p_{1/2}$   & & 2169.7& 8.8125  & -0.0004 	&-0.7028 	&0.0301 & -0.8323 & 7.9802 &\\ 
 
   \hspace{0.9cm}    $\rightarrow$ $7p_{3/2}$   & & 1833.1&  4.3391  & -0.0004	 &-0.3692& 	0.0183 & -0.4315 & 3.9077 &\\

 $5f_{7/2}  $    $\rightarrow$ $6p_{3/2}$   & 1026.8 & 1026.5

& -12.6148   & -0.0022 	&0.6548 & 0.1347&  0.8544 & -11.7603 &\\
   
   \hspace{0.9cm}    $\rightarrow$ $7p_{3/2}$   & & 1855.6&   10.6906  & -0.0009 	&-0.8479	 &0.0433&  -0.9884 & 9.7023 & \\
    
  $5g_{7/2}  $    $\rightarrow$ $5g_{9/2}$  & 8333333.3& 8333333.3&  -9.5003  & 0.0000	& 0.2725	& 0.0162 & 0.2981 & -9.2022 & \\  
    
 $6p_{1/2}$     $\rightarrow$ $6p_{3/2}$   & 5719.5& 5637.0

& 7.3339 &  0.0018	& -0.2834 &	-0.1021 &  -0.3899 & 6.9439 &\\
 
   \hspace{0.9cm}   $\rightarrow$ $7p_{3/2}$   & & 591.5&  -3.0526  & 0.0009	& 0.0444	 &-0.0437 & 0.0599 & -2.9927 & \\
 
$6p_{3/2}  $    $\rightarrow$ $7p_{1/2}$   &  &700.0 &  -4.8266  & 0.0008	& 0.0525 & -0.0354
&  0.0805 & -4.7461 &\\
 
 \hspace{0.9cm}   $\rightarrow$ $7p_{3/2}$   & & 660.9&  4.0170  & -0.0009	&   -0.0381	 &0.0411 & -0.0561 & 3.9608 &\\ 
 
 $7p_{1/2}  $   $\rightarrow$ $7p_{3/2}$   & & 11817.5&   -22.7635  & -0.0004&  0.9392	 &0.0434
&  1.0574 & -21.7061 & \\

 \hline 

\end{tabular}
\label{table:nonlin} 
\label{III}
\begin{tiny}
 \hspace{-14.9cm} $a \rightarrow$ Ref. \cite{Yoca2012}
\end{tiny}

\end{table}

The $M1$ transition amplitudes are presented in Table~\ref{IV} along with the different correlation  contributions. As expected, our results are consistent with the earlier calculations \cite{Dixit2007, Dutta2013, Roy2014, Safronovacan2011,  SafronovaLa2014} on these magnetic dipole transitions for other ionic species.  The amplitude of $5d_{3/2} \rightarrow 5d_{5/2}$ transition for W VI is available in literature using the HFR+CPOL method \cite{Yoca2012}, agreeing with the present result with 0.2\%.   The strong amplitudes between the fine structure states of same $^{2S+1}L$ level are dominated by the DF values and have correlation contributions of less than 0.1\%.  Therefore, the DF calculations for the $M1$ transitions between the fine structure states are excellent approximations to the total. 
  \begin{table*}[ht]

 \footnotesize
 \caption{ Calculated $M1$ transition amplitudes with the different correlation  contributing terms (in a.u.).   The 'Other$^a$' indicates the result obtained using
another method.}
\centering
\begin{tabular}{ cc  c ccc c c }

\hline \hline

 Transition    &   DF &  Core corr &  Pair corr  &  Core pol  & $\delta_\mathsf{corr}$   & Total & Other$^a$ \\ [0.2ex]
\hline

$5d_{3/2}  $   $\rightarrow$ $5d_{5/2}$   
& -1.54784 &  0.00789 &	0.00040	&  -0.00140
& -0.00139 & -1.54924& -1.55288 \\

 $5f_{5/2}  $   $\rightarrow$ $5f_{7/2}$    & -1.85149 & 0.00064	&  0.00015& 	-0.00030
& 0.00097 & -1.85051 &\\

  $5g_{7/2}  $    $\rightarrow$ $5g_{9/2}$   
 &  -2.10814&  0.00001	&  0.00000	& -0.00003
 & -0.00008 & -2.10822 &\\  
    
 $6p_{1/2}$     $\rightarrow$ $6p_{3/2}$  
& 1.14487 & -0.00047  &	-0.00056 & 	-0.00009
&  -0.00061 & 1.14426& \\

   \hspace{0.9cm}    $\rightarrow$ $7p_{3/2}$  & 0.08089 &  -0.00032	& -0.00296	&  0.00021&  0.00389&  0.08477 &\\
 
 $6p_{3/2}  $    $\rightarrow$ $7p_{1/2}$   
 &  -0.09400 & -0.00026	& -0.00893  &	-0.00036
 &  -0.00233 & -0.09633 &\\

 $7p_{1/2}  $    $\rightarrow$ $7p_{3/2}$   &   -1.14243	&0.00018 &	0.00022 &	0.00004 &0.00035&  -1.14208 &\\[0.2ex]

 \hline 

\end{tabular}
\label{table:nonlin} 
\label{IV}
\begin{tiny}
\textbf{Note:} The wavelength corresponding to each transition can be found in Table~\ref{III}.
\end{tiny}
\begin{tiny}
 \hspace{2.9cm} $a \rightarrow$ Ref. \cite{Yoca2012}
\end{tiny}
\end{table*}

 \begin{table}[ht]

\caption{ The lifetime (in s) of  $5d_{5/2}$ and $6s_{1/2}$ states. The 'Other$^a$' indicates the results obtained by other author.}
\centering 
\begin{tabular}{cccc}
\hline \hline
State &  & Our &Other$^a$  \\  [0.2ex]          
\hline

$5d_{5/2}$ &   & 1.40$\times 10^{-1}$ & 1.40$\times 10^{-1}$

\\ 
$6s_{1/2}$  &  & 3.82$\times 10^{-4}$& 3.77$\times 10^{-4}$ \\ [0.2ex]
 \hline

\end{tabular}
\label{table:nonlin}
\label{V}\\
\begin{tiny}
 \hspace{-2.9cm} $a \rightarrow$ Ref. \cite{Yoca2012}
\end{tiny}
\end{table}
The lifetimes of the first and second  excited  states are presented in Table~\ref{V}. Both the lifetimes are calculated using the RCC forbidden transitions ($E2$ and $M1$) amplitudes and corresponding experimental wavelengths obtained from the NIST. Our calculations show the metastable state $5d_{5/2}$ has lifetime of about 0.14 second which can be verified in the EBIT experiment \cite{Alfred,currell2003} therefore, W VI can be a good candidate for heavy ion storage ring \cite{Schippers2015}. Since the $E2$ and $M1$ matrix elements for the $5d_{5/2}-5d_{3/2}$ transition  have almost same order in magnitude, the lifetime of the $5d_{5/2}$ state is almost controlled by the $M1$ transition  (See Eq. (2.4) and Eq. (2.5) of Ref. \cite{Roy2014} and Eq.~\ref{3} of present work). Both the  HFR+CPOL \cite{Yoca2012} calculations of  lifetimes are well agreed with our corresponding results.  

 \begin{table}[ht]

\caption{ Hyperfine  $A$   constants  with different correlation contributing terms (in MHz). }
\centering
\begin{tabular}{ccccccc}

\hline \hline
  
  State        &  DF &  Core corr &  Pair corr &  Core pol &  $\delta_\mathsf{corr}$ &  RCC 
        \\ [0.2ex]
\hline

$5d_{3/2}  $ &  379.50 &   2.10 &  
20.33 &  	34.76 &  
 62.06 &   441.56   \\
 
$5d_{5/2}  $ &  147.14 &   1.30	&  5.66 & 	-96.28  & 
 -80.26 &   66.88  \\
 
 $5f_{5/2}  $ &  16.86 &  1.01 &  	3.50 &  	5.62 &  
  13.38  &  30.24   \\ 
 
$5f_{7/2}  $ &  9.42 &   0.47 &  	1.83 &  	-15.57 &  
 -13.60 &  -4.18   \\
   
$5g_{7/2}  $ &  0.58 &   0.00	 &  0.04 &  	0.06 &  
  -0.05 &   0.53   \\
   
   $5g_{9/2}  $&  0.37 & 0.00 &  	0.02 &  	-0.70 &  
 -0.67 &   -0.30  \\
   
$6s_{1/2}  $ &  9759.05 &  -158.12 &  	920.29 &  	860.81 &  
 1564.44 &   11323.50   \\
 
 $6p_{1/2}  $&   2343.47 &   -29.92 &  	265.56 &  	222.03 &  
 454.95 &   2798.42   \\
 
$6p_{3/2}  $ &  258.34 &   -1.59 &  	28.51 &  	53.67 &  
  87.77 &   346.11   \\
  
$7s_{1/2}  $  &  4243.79 &   -54.02 &  	220.91	&   336.70 & 
 498.22 &   4742.01   \\
 
 $7p_{1/2}  $ &  1131.38 &   -10.82 &  	79.28 &  	94.87 &  
 165.08 &   1296.46   \\
 
 $7p_{3/2}  $&   129.14 &    -0.44 &  	10.38	 &  23.73 &  
 41.62 &   170.76   \\
 
 $8s_{1/2}  $  &  2931.87 &   -33.73 &  	18.11	&   222.04 & 
203.73 &   3135.59   \\

 \hline 

\end{tabular}
\label{table:nonlin} 
\label{VI}
\end{table}

\begin{table}[ht]

\caption{ Hyperfine  $B$ constants  with different correlation
 contributing terms (in MHz). }
\centering
\begin{tabular}{ccccccc}

\hline \hline

  State         & DF &  Core corr &  Pair corr &  Core pol &    $\delta_\mathsf{corr}$ & RCC     \\ [0.2ex]
\hline

$5d_{3/2}  $  & -1958.31 & -38.74	&   -105.78	&   -287.29 &   
 -427.60 & -2385.91 \\
 
$5d_{5/2}  $  & -2321.05
& -50.07	&    -89.60	&    -477.65 &   
 -607.46 &  -2928.52 \\
 
 $5f_{5/2}  $  & -131.78
& -7.80	&   - 27.37	&    -372.74 &   
 -349.28 &  -481.06 \\ 
 
$5f_{7/2}  $  & -153.22     
& -8.86 &   	-29.93 &   	-445.18 &   
-419.46 &  -572.68 \\
   
$5g_{7/2}  $  & -5.66
& -0.08 &   	-0.38 &   	-213.43 &   
-206.36 &  -212.02 \\
   
   $5g_{9/2}  $ & -6.19 & -0.09	&   -0.41	&   -233.46 &   
 -225.87 & - 232.06 \\

$6p_{3/2}  $ 
& -3946.23 & 44.64	&  -434.43	&   -736.07 &   
-1113.48 & -5059.71 \\

 $7p_{3/2}  $ &
   
 -1972.79 &    14.65 &   	-158.56 &    	-315.91&  -464.47 & -2437.25 \\

 \hline 

\end{tabular}
\label{table:nonlin} 
\label{VII}
\end{table}

The hyperfine structure constants $A$ and $B$  of W VI with mass number 183 and nuclear spin 1/2, impotant parameters for laboratory plasma and EPR property of molecules, are presented in Tables ~\ref{VI} and ~\ref{VII}, respectively. Also, these constants are important contributors for high resolution spectroscopy. Both these constants are presented with different correlation contributions. The hyperfine $A$ values have correlation contributions within 7\% to 34\% except for the states $5d_{5/2}$, $5f_{5/2}$, $5f_{7/2}$ and $5g_{9/2}$ where contributions are 55$\%$, 79$\%$, 144$\%$ and 181$\%$, respectively. Here in most of the cases, the core polarization and pair correlation contributions are comparable in magnitudes except in the cases of $5d_{5/2}$, $5f_{7/2}$ and $5g_{9/2}$ states, where contributions for the former correlation term are one order more than latter term.   As seen from the Table~\ref{VII}, the $B$ constants of $5f_{5/2}$, $5f_{7/2}$,  $5g_{7/2}$ and $5g_{9/2}$ states are  correlated abnormally. Similar kind of features are  observed in many other systems, like, Ga III \cite{Dutta2013}, In III \cite{Roy2014},  Sc III \cite{Dutta2015} where core polarization is much larger than DF values. For other states, $B$ constants are correlated by about 22\% to 28\%. Using these two hyperfine parameters, the hyperfine energy shift of a hyperfine level $F$ of a low-lying state due to nuclear spin $I$  can be calculated easily using Eq.~(\ref{4}).

\begin{table*}[!]

\caption{ Hyperfine  $A$   constants (in MHz) for different isotopes of W VI. The parenthesis indicate mass numbers of the  isotopes. }
\centering
\begin{tabular}{ccccc}

\hline \hline
  
  State        &  $A (182)$ &  $A (183)$ &  $A (184)$  &  $A (186)$  
        \\ [0.2ex]
\hline

$5d_{3/2}  $ &  488.30 &   441.56 &  
541.70 &  	 576.36    \\
 
$5d_{5/2}  $ &  73.94 &  66.88	&  82.05 & 	87.33   \\
 
 $5f_{5/2}  $ &  33.44 &  30.24  &  	37.09 &  	39.46  \\ 
 
$5f_{7/2}  $ &  -4.62 &   -4.18 &  	 -5.12 &  	-5.44   \\
   
$5g_{7/2}  $ &  0.59 &  0.53	 &  0.65 &  	0.70   \\
   
   $5g_{9/2}  $&  -0.33 & -0.30 &  	-0.37 &  	-0.39 \\
   
$6s_{1/2}  $ &  12522.97 &  11323.50 &  	13890.60 &  	14777.21 \\
 
 $6p_{1/2}  $&   3094.63 &   2798.42 &  	 3433.08 &  	3652.71 \\
 
$6p_{3/2}  $ & 382.74  &   346.11 &  	  424.61    &  	451.78   \\
  
$7s_{1/2}  $  &   5244.32 &   4742.01 &  	5817.06	&   6188.35 \\
 
 $7p_{1/2}  $ & 1433.69  &   1296.46 &  	1590.49 &  	1692.24    \\
 
 $7p_{3/2}  $&  188.83  &    170.76 &  209.48		 &  222.89 \\
 
 $8s_{1/2}  $  &   3467.74 &  3135.59 	 &  3846.46 &  	 4091.98 \\

 \hline 

\end{tabular}
\label{table:nonlin} 
\label{VIII}
\end{table*}

In Table~\ref{VIII} and ~\ref{IX}, we have shown the hyperfine structure constants for the different stable isotopes of W VI.  We  consider the isotopes having mass number 182, 183, 184 and 186 with corresponding  nuclear spin values 2, 0.5, 4 and 2, respectively.
Results in these tables are important for line-profile measurements of absorption and emission lines of W VI in different astronomical bodies, and hence to get accurate picture of abundances of tungsten.
\begin{table}[ht]

\caption{ Hyperfine  $B$ constants  (in MHz) for different isotopes of W VI. The parenthesis indicate mass numbers of the  isotopes.}
\centering
\begin{tabular}{ccccc}

\hline \hline

  State         &$B (182)$ &  $B (183)$ &  $B (184)$  &  $B (186)$      \\ [0.2ex]
\hline

$5d_{3/2}  $  &  -2871.18 & -2385.91	&   -2520.71	&   -2116.31  \\
 
$5d_{5/2}  $  & -3524.15
& -2928.52&   -3093.96	&    -2597.60  \\
 
 $5f_{5/2}  $  &  -578.90
&  -481.06	&     -508.23	&    -426.70  \\ 
 
$5f_{7/2}  $  & -689.17    
& -572.68 &   	-605.04 &   	-507.97   \\
   
$5g_{7/2}  $  & -255.15
& -212.02 &   	 -224.00 &   	-188.06   \\
   
   $5g_{9/2}  $ & -279.26 & -232.06	&   -245.17	&   -205.84  \\

$6p_{3/2}  $ 
& -6088.80 & -5059.71	&  -5345.56 	&    -4487.99  \\

 $7p_{3/2}  $ &-2932.97
   
  &   -2437.25  & -2574.95  	 &    	-2161.86 \\

 \hline 

\end{tabular}
\label{table:nonlin} 
\label{IX}
\end{table}

The theoretical uncertainties in the calculated property parameters can be  estimated by  the quality of the wave functions,  especially where the amplitudes are significants,  at the levels of the Dirac-Fock. Along with that we should consider the contributions from the other correlation terms  than those considered in this paper and  quantum electro-dynamic effects  (totally at most ~$\pm 2\%$). As a result, our estimated  maximum uncertainties are $\pm 3\%$ for the $E1$ amplitudes $\pm 6\%$ for forbidden transitions and hyperfine constants. 
\section{CONCLUSION}

We have calculated the transition amplitudes of allowed  and forbidden  transitions for W VI using a highly correlated relativistic coupled-cluster theory.  Hyperfine structure $A$ and $B$ constants of few low-lying states for the various isotopes of this element
are estimated where comparison could not be done due to lack of  theoretical or experimental endeavour. Importance of consideration of pair correlation in the many-body approaches are studied along with detail of core polarization correlation contributions. Good agreements are achieved between the electric dipole matrix elements based on the length gauge and velocity gauge. Our correlation exhaustive many-body approach provide  scope to experimentalists  to test their up to date technologies.     The forbidden infra-red  and optical transitions among the fine structure levels of the $5d$ and $6p$ terms, respectively, are very important for laser spectroscopy, plasma research and different atomic physics experiments. Our spectroscopic estimations of these allowed and forbidden transition lines mitigate the demand of high resolution data observed from stellar and interstellar medium. Our hyperfine data for various isotopes are also supplement to this.

\section*{ACKNOWLEDGMENTS}
The calculations carried out in the IBM cluster at IIT-Kharagpur, India funded by DST-FIST (SR/FST/PSII-022/2010).

\clearpage

\clearpage




\begin{thebibliography}{}


\bibitem{Clementson2015}
 Clementson J,  Lennartsson T and  Beiersdorfer P 2015  Atoms \textbf{3} 407


\bibitem{Riccardo2009}
 Riccardo V,   Firdaouss M,  Joffrin E,  Matthews G,  Mertens P,  Thompson V and  Villedieu E 2009  Phys. Scr. T, \textbf{138} 014033
\bibitem{Rohde2009}
 Rohde V,   Balden M,  Lunt T and the ASDEX Upgrade Team, 2009 Phys. Scr. \textbf{2009} 014024
\bibitem{Reinke2010}
Reinke M,  Beiersdorfer P,  Howard N T,  Magee  E W,   Podpaly Y,  Rice J E- and  Terry J L 2010  Rev. Sci. Instrum. \textbf{81} 10D736

\bibitem{Rakhimov2000}
 Rakhimov R R,  Jones D E,  Rocha H L,  Prokof'ev A I and  Aleksandrov A I 2000 J. Phys. Chem. B, \textbf{104} 10973
\bibitem{Mei2016}
Mei Y,  Wei C -F,  Zheng W -C 2016 Physica B: Condensed Matter Volume \textbf{483} 78

 \bibitem{Miyahara1991}
 Miyahara A and  Behrisch R 1991 Journal of Nuclear Materials, \textbf{179} 1231
\bibitem{Behrisch1991}
 Behrisch R 1991 Atomic and Plasma-Material Interaction Data for Fusion  \textbf{1} 7
\bibitem{Armstrong2013}
 Armstrong D E J,   Yi X,   Marquis E A and   Roberts S G 2013 Journal of Nuclear Materials \textbf{432} 428 

\bibitem{Safronova2014}
 Safronova M S,  Dzuba V A,  Flambaum V V,  Safronova U I,  Porsev S G and  Kozlov M G 2014 Phys. Rev. Lett. \textbf{113} 030801 
\bibitem{Alfred}
 M\"uller A 2015 Atoms \textbf{3} 120
\bibitem{Seaton1954}
 Seaton M J 1954  Mon. Not. R. Astron. Soc. \textbf{114} 154 
\bibitem{Seaton1957}
Seaton M J and  Osterbrock D E 1957  Astrophys. J. \textbf{125} 66



\bibitem{Biemont1996}
Biemont E  and  Zeippen C J 1996 Comment. At. Mol. Phys.
\textbf{33} 29


\bibitem{Lee2002}
 Lee D -C,  Halliday A N,  Leya I,  Wieler R,  Wiechert U 2002 Earth and Planetary Science Letters \textbf{198} 267 
\bibitem{Masarik1997}
Masarik J 1997 Earth and Planetary Science Letters \textbf{152} 181

\bibitem{Casassus2005}
Casassus S,  Storey P J,  Barlow M J and  Roche P F 2005 MNRAS, \textbf{359} 1386
\bibitem{Podpaly2009}
Podpaly Y,  Clementson J,  Beiersdorfer P,  Williamson J,  Brown G V and Gu M F 2009 Phys. Rev. A \textbf{80}(5) 052504 
\bibitem{Clementson2010}
Clementson J,  Beiersdorfer P and  Gu M F 2010
Phys. Rev. A \textbf{81}(1) 012505
\bibitem{Koutsospyros2006}
 Koutsospyros A,  Braida W,  Christodoulatos C, Dermatas D, Strigul N 2006  J.Hazardous Materials, \textbf{136} 1 

\bibitem{Schmitz1992}
 Schmitz R A,  Albracht S P J,  Thauer R K 1992  FEBS Letters, \textbf{309} 78


\bibitem{Meijer1974}
 Meijer F G 1974 Physica  \textbf{73} 415

\bibitem{Sugar1975}
Sugar J and  Kaufman V 1975 Phys. Rev. A \textbf{12} 994
\bibitem{Kramida2006}
Kramida A E and  Reader J 2006 At. Data Nucl. Data Tables \textbf{92} 457
\bibitem{Loch2006}
Loch S D,   Pindzola M S,  Ballance C P,  Griffin D C,   Whiteford A D and  P\"{u}tterich T 2006 AIP Proc. \textbf{874} 233
\bibitem{Kramida2009}
Kramida A E and  Shirai T 2009 At. Data Nucl. Data Tables \textbf{95} 305
\bibitem{Safronova2009}
 Safronova U I,  Safronova A S and  Beiersdorfer P 2009 J. Phys. B: At. Mol. Opt. Phys. \textbf{42} 165010
\bibitem{Safronova2011}
Safronova U I,  Safronova A S,  Beiersdorfer P and Johnson W R 2011 J. Phys. B: At. Mol. Opt. Phys. \textbf{44} 035005
\bibitem{Safronova2012a}
Safronova U I,  Safronova A S and  Beiersdorfer P 2012
J. Phys. B \textbf{45} 085001
\bibitem{Safronova2012b}
Safronova U I and  Safronova A S 2012 Phys. Rev. A \textbf{85} 032507 

\bibitem{Seidel2009}
Biedermann C,  Radtke R,  Seidel R and  Behar E 2009 J. Phys. Conf. Series \textbf{163} 012034
\bibitem{Dalhed1986}
Dalhed S,  Nilsen J and  Hagelstein P 1986 Phys. Rev. A \textbf{33} 264 
\bibitem{Behar1997}
Behar E,  Peleg A,  Doron R,  Mandelbaum P and  Schwob J L 1997 J. Quant. Spectrosc. Radiat. Transf. \textbf{58} 449
\bibitem{Behar1999}
Behar E,  Mandelbaum P and  Schwob J L 1999 Phys. Rev. A \textbf{59} 2787
\bibitem{Behar2000}
Behar E,  Doron R,  Mandelbaum P and  Schwob J L 2000 Phys. Rev. A  \textbf{61} 062708
\bibitem{Clementson2008}
Clementson J,   Beiersdorfer P,  Gu M F,  McLean H S and  Wood R D 2008  J. Phys. Conf. Ser. \textbf{130} 001204
\bibitem{Meng2008}
 Meng F -C,  Chen C -Y,  Wang Y -S and  Zou Y -M 2008 J. Quant. Spectrosc. Radiat. Transf. \textbf{109} 2000 


\bibitem{Biedermann2009}
Biedermann C,  Radtke R,  Seidel R and  P\"{u}tterich T 2009  Phys. Scr. T \textbf{134} 014026
\bibitem{Meng2009}
 Meng F -C,  Zhou L, Huang M,  Chen C -Y,  Wang Y -S and  Zou Y -M 2009 J. Phys. B: At. Mol. Opt. Phys. \textbf{42} 105203 
\bibitem{Radtke2009}
 Biedermann C and  Radtke R 2009 AIP Conf. Proc. \textbf{1161} 95 

\bibitem{Ballance2010}
 Ballance C P,  Loch S D,  Pindzola M S and  Griffin D C 2010 J. Phys. B: At. Mol. Opt. Phys. \textbf{43} 205201 

\bibitem{Schippers2011}
 Schippers S,  Bernhardt D,  M\"{u}ller A,  Krantz C,  Grieser M,  Repnow R,  Wolf A,  Lestinsky M,  Hahn M 2011 \textit{et al.} Phys. Rev. A \textbf{83} 012711
\bibitem{Yoca2012}
 Yoca S E,  Palmeri P,  Quinet P,  Jumet G and  Bi\'{e}mont \'{E} 2012 J. Phys. B: At. Mol. Opt. Phys. \textbf{45}  035002 
\bibitem{Migdalek2014}
 Migdalek J and  Siegel W 2014 J. Phys. B: At. Mol. Opt. Phys. \textbf{47} 075003 
\bibitem{Dixit2007}
Dixit G,  Sahoo B K,   Chaudhuri R K and
 Majumder S 2007 Phys. Rev. A \textbf{76} 042505
\bibitem{Dixit2008}
 Dixit G,  Nataraj H S,  Sahoo B K,   Chaudhuri R K and
 Majumder S 2008 Phys. Rev. A \textbf{77} 012718 
\bibitem{Dutta2016}
 Dutta N N and  Majumder S 2016 Indian J. Phys., \textbf{90} 373



\bibitem{Urban1985}
 Urban M,  Noga J,  Cole S J and  Bartlett R J 1985 J. Chem. Phys. \textbf{83} 4041 
\bibitem{Bishop1987}
Bishop R F  and  K\"{u}mmel H G 1987 Physics Today March {\bf 40}, 52

\bibitem{Lindgren1987}
 Lindgren I  and  Mukherjee D 1987 Physics Reports \textbf{151} 93


\bibitem{Raghavachari1989}
 Raghavachari K,  Trucks G W,  Pople J A and Head-Gordon M 1989 Chem. Phys. Lett. {\bf 157} 479
\bibitem{Ilyabaev199293}
Ilyabaev E and  Kaldor U  1992 Chem. Phys. Lett. {\bf 194} 95, 1993 Phys. Rev. A {\bf 47} 137
\bibitem{Chaudhuri2003}
Chaudhuri R K,  Sahoo B K,  Das B P,  Merlitz H,  Mahapatra U S 2003 J. Chem. Phys. {\bf 119} 10633
\bibitem{Mani2009}
 Mani B K,  Latha K V P and  Angom D  2009 Phys. Rev. A {\bf 80} 062505
\bibitem{Mani2011}
 Mani B K  and  Angom D 2011 Phys. Rev. A {\bf 83} 012501
\bibitem{Dutta2011}
 Dutta N N  and  Majumder S 2011 Astrophys. J., {\bf 737} 25 

\bibitem{Dutta2013}
 Dutta N N,  Roy S,  Dixit G  and  Majumder S 2013
 Phys. Rev. A \textbf{87} 012501 


\bibitem{Roy2014}
 Roy S,  Dutta N N  and  Majumder S 2014   Phys. Rev. A \textbf{89} 042511
\bibitem{Dutta2014}
 Dutta N N and  Majumder S 2014 Phys. Rev. A  {\bf 90} 012522 
\bibitem{Roy2015}
 Roy S and  Majumder S 2015  Phys. Rev. A \textbf{92} 012508 

\bibitem{Dutta2015}
 Dutta N N,  Roy S and  Deshmukh P C 2015 Phys. Rev. A \textbf{92} 052510
\bibitem{Bhowmik2017}
Bhowmik A,  Dutta N N and  Roy S 2017 Astrophys. J., {\bf 836} 125
\bibitem{Lindgren1985}
Lindgren I and  Morrison J  \textit{Atomic Many-body Theory}, edited by  Ecker G,  Lambropoulos P and  Walther H (Springer, Berlin, 1985), Vol. \textbf{3}.
\bibitem{Pasteka2017}
Pa$\check{\textrm{s}}$teka L F,  Eliav E,  Borschevsky A,  Kaldor U and  Schwerdtfeger P 2017 Phys. Rev. Lett. \textbf{118} 023002
\bibitem{Bishop1991}
Bishop R F 1991 Theor. Them. Acta  \textbf{80} 95 

\bibitem{Mondal2013}
 Mondal P K,  Dutta N N,  Dixit G and  Majumder S 2013  Phys. Rev. A \textbf{87} 062502
\bibitem{Grant1974}
 Grant I P 1974 J. Phys. B \textbf{7} 1458
\bibitem{Johnson1995}
Johnson W R,  Plante D R and  Sapirstein J 1995 Adv. At. Mol. Opt. Phys. \textbf{35} 255 
 

\bibitem{Cheng1985}
 Cheng K T and  Childs W J 1985 Phys. Rev. A \textbf{31} 2775 
 \bibitem{Visscher1997}
Visscher L and  Dyall K G 1997 Atomic Data and Nuclear Data Tables \textbf{67} 207


\bibitem{Motecc1990}
 Clementi E (Ed.), \textit{Modern Techniques in Computational Chemistry: MOTECC-90}, (ESCOM
Science Publishers B. V., 1990).
\bibitem{Parpia2006}
 Parpia F A,  Fischer C F and  Grant I P 2006  Comput. Phys. Commun. \textbf{175} 745 
 \bibitem{Gaunt1929}
 Gaunt J A 1929 Proceedings of the Royal Society A \textbf{122} 513
 \bibitem{Mann1971} 
 Mann J B and  Johnson W R 1971 Phys. Rev. A \textbf{4}, 41 
\bibitem{NIST2015}
 Kramida A,  Ralchenko Y,   Reader J and NIST ASD Team (2015). NIST Atomic Spectra Database (ver. 5.3),  http://physics.nist.gov/asd [2015, December 27]. National Institute of Standards and Technology, Gaithersburg, MD.
\bibitem{Grant2007}
 Grant I P, {\it Relativistic Quantum Theory of Atoms and Molecules: Theory and
Computation}, (Springer, 2007).
\bibitem{Safronovacan2011}
Safronova U I and  Safronova M S 2011 Can. J. Phys. {\bf 89} 465 
\bibitem{SafronovaLa2014}
 Safronova U I and  Safronova M S  2014 Phys. Rev. A {\bf 89} 052515
\bibitem{currell2003}
 Currell F J 'Electron Beam Ion Traps and their Use in the Study of Highly Charged Ions'. In 'The Physics of Multiply and Highly Charged Ions', {\bf 1}, 39, Currell, F., Ed.; Kluwer Academic
Publishers: Dordrecht, The Netherlands, (2003).
\bibitem{Schippers2015}
 Schippers S 2015 Nucl. Instrum. Meth. in Phys. Res. \textbf{350} 61





\end{thebibliography}

\end{document}